\documentclass[11pt,twoside]{article}
\usepackage{amsfonts}

\usepackage[dvips]{graphicx}

\input{tcilatex}

\begin{document}

\title{Plane 5D worlds and simple compactification}
\author{David Solano$^{1,2}$ and Rodrigo Alvarado$^{1}$ \thanks{%
email: dsolano@efis.ucr.ac.cr, dsscr@yahoo.com; realmar@cariari.ucr.ac.cr} \\
$^{1}$Escuela de Fisica, Universidad de Costa Rica\\
Cuidad Universitaria Rodrigo Facio, Montes de Oca, San Jose.\\
and\\
$^{2}$Laboratorio Nacional de Nanotecnologia, \ \\
Centro Nacional de Alta Tecnologia. \\
1 km south from the U.S. Embassy \\
Pavas. San Jose, Costa Rica}
\maketitle

\begin{abstract}
We obtain a new kind of exact solution to vaccum Einstein field equations
that contain both Minkowsian world and a special 5D curved spacetime with
particularly free structure. This special world is defined by an arbitrary
function and a space of three parameters. We suggest that this solution
could correct (in principle) certain aspects of the physics in flat
spacetime.
\end{abstract}

\newpage

\section{Introduction}

The use of a fifth dimension in Theoretical Physics has been a common way to
accomplish the unification of the gravitation and the other interactions
since the seminal work of Nordstr\"{o}m \cite{Nordstrom} and Kaluza \cite
{Kaluza}. \ The problem of recover the standard 4-D spacetime can be solve
by the Klein's mecanism\cite{Klein}, which explains how the additional
coordinates are length-like and could be detected. The idea of
extra-dimension compactification is necessary, there are also certain
research oriented to the idea extra-coordinates are not physically real (P%
\textit{rojective theories}) \cite{schmutzer}. There are also the \textit{%
Non-compactified theories}, in which extra-dimensions are not necessarily
length-like or compact (see \cite{Overduin} and references therein).

In recent years, higher dimensional theories of extended objects, like
String Theory \cite{polchinski}, provide a promising scheme to construct
renormalizable quantum fields that unify all the interactions of Nature.

Since the exact solution found by Davidson and Owen in the mid-eighties \cite
{owen}, the dynamics of 5D-physics has been widely study since the late
eighties by Wesson, and his collaborators in their Theory of Induced Matter 
\cite{wesson}. Latter, Randall and Sundrum \cite{randall} proused a
Brane-theory in 5D that explain the large hierarchy between the quantum
gravity effect and the standard model. In \cite{brax} there an extensive
review full of reference of Randall-Sundum Brane worlds principally applied
to cosmology.

In the present work, the existence of solutions to the Einstein's field
equations when the metric tensor dependents only of a new extra-coordinate
in plane symmetry are discussed. The anzatz followed in the present article
is mainly based on \cite{alvarado}, who showed that the compactification
(vanishing) of one coordinate could happen for vacuum gravity for a static
cylindrical world. The philosophy in the work is to discard a wide range of
solutions because they do not represent asimptotically flat spacetimes. But
we show in \cite{david} that a large family of spacetimes can be constructed
by using an internal freedom that Einstein's field equation give for this
specific symmetry and the Israel's junction conditions.

\section{Field equation and its solutions}

Let be the following line element:

\begin{equation}
ds^{2}=e^{\gamma (\eta )}dt^{2}-e^{\tau (\eta )}dx^{2}-e^{\mu (\eta
)}dy^{2}-e^{\nu (\eta )}dz^{2}-e^{\rho (\eta )}d\eta ^{2}  \label{metric}
\end{equation}

where the new coordinate '' $\eta $ '' is introduced. Here, all the
functions $\gamma ,\tau ,\mu ,\nu ,\rho $ are only $\eta $-dependent. Notice
that our procedure is similar in spirit to that of \cite{vilenkin}, only
that we do not include a thin matter wall in energy-stress tensor
proportional to a Dirac $\delta $-function.

Let us define the symbol $()^{\prime }$ as $d/d\eta $ . Then, the vaccum
Einstein field equations ($R_{AB}=0$) are

\begin{eqnarray}
\gamma ^{\prime \prime }+\gamma ^{\prime }{}^{2}-\gamma ^{\prime }\rho
^{\prime }+\gamma ^{\prime }(\tau ^{\prime }+\mu ^{\prime }+\nu ^{\prime })
&=&0  \label{g_t} \\
\tau ^{\prime \prime }+\tau ^{\prime }{}^{2}-\tau ^{\prime }\rho ^{\prime
}+\tau ^{\prime }(\gamma ^{\prime }+\mu ^{\prime }+\nu ^{\prime }) &=&0
\label{g_x} \\
\mu ^{\prime \prime }+\mu ^{\prime }{}^{2}-\mu ^{\prime }\rho ^{\prime }+\mu
^{\prime }(\gamma ^{\prime }+\tau ^{\prime }+\nu ^{\prime }) &=&0
\label{g_y} \\
\nu ^{\prime \prime }+\nu ^{\prime }{}^{2}-\nu ^{\prime }\rho ^{\prime }+\nu
^{\prime }(\gamma ^{\prime }+\tau ^{\prime }+\mu ^{\prime }) &=&0
\label{g_z} \\
2({\gamma }^{\prime \prime }+{\tau }^{\prime \prime }+{\mu }^{\prime \prime
}+{\nu }^{\prime \prime })-{\rho }^{\prime }({\gamma }^{\prime }+{\tau }%
^{\prime }+{\mu }^{\prime }+{\nu }^{\prime })+{\gamma }^{\prime 2}+{\tau }%
^{\prime 2}+{\mu }^{\prime 2}+{\nu }^{\prime 2} &=&{0}  \label{g_eta}
\end{eqnarray}

Let $\chi =\gamma +\tau +\mu +\nu $, an arbitrary function where its square
is naturally

\begin{equation}
\chi ^{\prime 2}=\gamma ^{\prime 2}+\tau ^{\prime 2}+\mu ^{\prime 2}+\nu
^{\prime 2}+2\gamma ^{\prime }(\tau ^{\prime }+\mu ^{\prime }+\nu ^{\prime
})+2\tau ^{\prime }(\mu ^{\prime }+\nu ^{\prime })+2\mu ^{\prime }\nu
^{\prime }
\end{equation}

And now by summing (\ref{g_t}), (\ref{g_x}), (\ref{g_y}) and (\ref{g_z}) we
get

\begin{equation}
2\chi ^{\prime \prime }+\chi ^{\prime }{}^{2}-\rho ^{\prime }\chi ^{\prime
}=0  \label{solution_1}
\end{equation}

which gives the integral

\begin{equation}
\rho =\xi +2\ln (\chi ^{\prime })+\chi  \label{condition}
\end{equation}

where $\xi $ is a arbitrary constant. The value of $\xi $ can be perfectly
chosen as zero, because in $g_{\eta \eta }=e^{\rho }=\chi ^{\prime
}{}^{2}e^{\chi }e^{\xi }$ the constant $e^{\xi }$ is just an amplification
parameter of the length element that only depends of the chosen unit system.
According to (\ref{solution_1}) and (\ref{g_t}) it is easy to see that:

\begin{equation}
\gamma =A_{0}\chi +B_{0}  \label{gamma}
\end{equation}

Applying this procedure to equations (\ref{g_x}), (\ref{g_y}) and (\ref{g_z}%
), we get

\begin{eqnarray}
\tau &=&A_{1}\chi +B_{1}  \label{tau} \\
\mu &=&A_{2}\chi +B_{2}  \label{mu} \\
\nu &=&A_{3}\chi +B_{3}  \label{nu}
\end{eqnarray}

Without loss of generally, let us choose $B_{\alpha }=0,$($\vee \alpha
=0,1,2,3$). Therefore, every metric function (except that one related to the
extra-coordinate) maintain a linear relation with $\chi $. \ Finally, by
using the relation in (\ref{solution_1}) in (\ref{g_eta}) we obtain the
important relation:

\begin{equation}
\chi ^{\prime }(F-1)=0  \label{solution_2}
\end{equation}

where $F=A_{0}^{2}+A_{1}^{2}+A_{2}^{2}+A_{3}^{2}$ is a real non-negative
constant. Therefore, from (\ref{solution_2}) we must analyze the two
separate cases:

\subparagraph{Case: $F\neq 1$}

This implies that necessary $\chi =$constant. \ Hence, $\chi $ is only a
scaling parameter, the 4-D Minkowskian space-time is recovered as we
expected if $\chi =$ $0$:

\begin{equation}
ds^{2}=dt^{2}-(dx^{2}+dy^{2}+dz^{2})
\end{equation}

\subparagraph{Case: \textbf{$F=1$ }}

According to equation (\ref{solution_2}), the $F$ parameter must be $1$, or

\begin{equation}
A_{0}^{2}+A_{1}^{2}+A_{2}^{2}+A_{3}^{2}=1  \label{aes_1}
\end{equation}

Since $\chi =\gamma +\tau +\mu +\nu $ by definition, it is clearly seen from
equations (\ref{gamma}) to (\ref{nu}) that

\begin{equation}
A_{0}+A_{1}+A_{2}+A_{3}=1  \label{aes_2}
\end{equation}
\qquad\ 

\begin{figure}[tbp]
\caption{The domain of $Q$: All the $\left( u,v\right) $ points on the
ellypse and inside it define the parametrization of the integration
constants }
\label{drawing}
\begin{center}
\includegraphics[width=4cm]{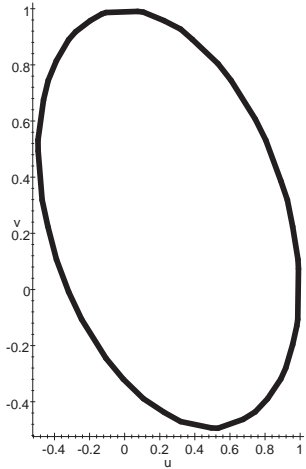}
\end{center}
\end{figure}

Then, one can find a map $Q:\Re^{2}\rightarrow \Re^{4}$ that describes the
solution to the algebraic equations (\ref{aes_1}) and (\ref{aes_2}). For
simplicity, let define $A_{0}=u$ and $A_{1}=v$ and then, the $Q $ map that
relates $(u,v)$ with $\left( A_{0},A_{1},A_{2},A_{3}\right) $ is given by
the parametrization

\begin{equation}
A=\left( 
\begin{array}{c}
A_{0} \\ 
A_{1} \\ 
A_{2} \\ 
A_{3}
\end{array}
\right) =\left( 
\begin{array}{c}
u \\ 
v \\ 
-\frac{1}{2}\left( u+v-1+\sqrt{\Delta }\right) \\ 
\frac{1}{2}\left( u+v+1+\sqrt{\Delta }\right)
\end{array}
\right)  \label{paremeter}
\end{equation}

where $\Delta =-3(u^{2}+v^{2})+2(u+v-uv)+1$. The necessity of real
integration constants restricts the domain of $Q$ to the region of the $uv$%
-plane described by the ellipse: $\Delta \left( u,v\right)
=-3(u^{2}+v^{2})+2(u+v-uv)+1=0$ and all its interior points (Figure \ref
{drawing}).

One particularly simple solution is, for example: $A_{0}=-\frac{1}{2}%
,\;A_{1}=$ $A_{2}=A_{3}=\frac{1}{2}$. Then, the line element under this
conditions would be

\begin{equation}
ds^{2}=e^{-\frac{1}{2}\chi }dt^{2}-e^{\frac{1}{2}\chi }dx^{2}-e^{\frac{1}{2}%
\chi }dy^{2}-e^{\frac{1}{2}\chi }dz^{2}-(\chi ^{\prime })^{2}e^{\chi }d\eta
^{2}  \label{solution_3}
\end{equation}

A formal classification of the spacetimes based on the behavior of the $\chi 
$-function will be treated with exhaustive details in \cite{david}.

\section{Conclussions}

Let summarize these results here. We found a large class of exact solution
to the field equation of Einstein gravity and discovered a new kind of
freedom (maybe of mathematical nature) that they permit. A spacetime with
the form \ref{solution_3} can be used to study quantum fields theories
(QFT's) is a curved spacetime that corrects the minkowkian plane world.
Depending on the form of the arbitrary $\chi $, we can construct a new
special classically fluctuting spacetime that can correct (in principle)
classical and quantum trayectories of particles \cite{david}. In a more
traditional stand, when we applied the conventional Kaluza-Klein unification
of electromagnetism and gravitation, EM energy ``fusses'' with the
gravitational one (that is locally indefined), in such way that Kaluza-Klein
world could contain zero total energy \cite{david2}.

We can call this non-trivial solution to Einstein equations \textbf{R0X}.
The \textit{R0} is for Ricci flatness that every vaccum solution has. The
``X'' is for a specific kind of the $\chi $- function we choose for a
spacetime. As we will discuss with extensive details in \cite{david2}, if $%
\chi $ is a map $\chi :\Bbb{V\rightarrow W}$ , where $\Bbb{V},\Bbb{W\subset R%
}$ and the target set $\Bbb{W}$ is finite ( $\Bbb{W}=[0,a]$ , $a\in \Bbb{R}$%
) and $\lim_{\eta \rightarrow \pm \infty }\chi =0$, then it is easy to see
from equation (\ref{solution_3}) \ that traditional flat world in 4D is
recovered. Thus, when $\chi $ is a well-tempered function the
extra-dimension disappears. We may suggest that this simple compactification
procedure (far more easier that the Calabi-Yau one) can be use to formulate
a consistent Brane theory that explain the large difference between the
electro-weak and Planck scale.

Now, if the domain $\Bbb{V}=\Bbb{R}-\{S_{n}\}$, where $\{S_{n}\}$ is a
finite set of point where $\chi $ is ill-defined, the R0X manifold would
have a finite number of point where the Riemann tensor does not exist and
thus the gravitational tidal forces would be infinite. As in \cite{david2}
we show that under certain conditions R0X has exact solutions to the
geodesic equations, R0X could be a ideal ``laboratory'' to explore certain
propousal made in the past, such as the Penrose's cosmic censorship
conjucture \cite{penrose}. Also, the study of new singulaties could
generated a rich arena for the recently developed Loop Quantum Gravity
theory \cite{lqg}.

\textbf{Acknowledgment.} D S would like to thank Jorge A. Diaz and Ms. Eda
Maria Arce for their kindness and hospitality at LANOTECH.

\end{document}